\documentclass[reprint]{revtex4-1}

\usepackage{amsmath}    
\usepackage{graphicx}   
\usepackage{verbatim}   
\usepackage{color}      
\usepackage{hyperref}   

\usepackage{multirow}

\begin{document}

\title{Asymmetric magnetic proximity interactions in MoSe$_{2}$/CrBr$_{3}$ van der Waals heterostructures}

\author{Junho Choi$^{1}$, Christopher Lane$^{2}$, Jian-Xin Zhu$^{2}$, Scott A. Crooker$^{1}$}
\affiliation{$^1$National High Magnetic Field Laboratory, Los Alamos National Laboratory, Los Alamos, New Mexico 87545, USA}
\affiliation{$^2$Theoretical Division and Center for Integrated Nanotechnologies, Los Alamos National Laboratory, Los Alamos, New Mexico 87545, USA}


\begin{abstract}

\end{abstract}

\maketitle
\textbf{Magnetic proximity interactions (MPIs) between atomically-thin semiconductors and two-dimensional magnets provide a means to manipulate spin and valley degrees of freedom in nonmagnetic monolayers, without the use of applied magnetic fields \cite{Zutic:2019, Gilbertini:2019, Mak:2019}.  In such van der Waals (vdW) heterostructures, MPIs originate in the nanometer-scale coupling between the spin-dependent electronic wavefunctions in the two materials, and typically their overall effect is regarded as an effective magnetic field acting on the  semiconductor monolayer \cite{Zhao:NatNano:2017, Zhong:SciAdv:2017, Norden:NatComm:2019, Zhong:NatNano:2020, Ciorciaro:PRL:2020}. Here we demonstrate that this picture, while appealing, is incomplete: The effects of MPIs in vdW heterostructures can be markedly asymmetric, in contrast to that from an applied magnetic field.  Valley-resolved optical reflection spectroscopy of MoSe$_{2}$/CrBr$_{3}$ vdW structures reveals strikingly different energy shifts in the $K$ and $K'$ valleys of the MoSe$_2$, due to ferromagnetism in the CrBr$_3$ layer. Strong asymmetry is observed at both the A- and B-exciton resonances. Density-functional calculations indicate that valley-asymmetric MPIs depend sensitively on the spin-dependent hybridization of overlapping bands, and as such are likely a general feature of such hybrid vdW structures. These studies suggest routes to selectively control \textit{specific} spin and valley states in monolayer semiconductors \cite{Xiao:PRL:2012, Xu:2014}.}

The ability of short-range proximity interactions to imbue magnetic functionality into otherwise nonmagnetic materials \cite{Lazic:2016, Kawakami:2018} has exciting prospects for devices that combine the optical and electrical properties of monolayer semiconductors \cite{MakShan:2016, Urbaszek:2018, Strano:2012} with additional tuning parameters that couple directly to spin and valley pseudospin \cite{Schaibley:2016}. The atomically-smooth surfaces that are nowadays routinely achieved with van der Waals materials allow for nearly ideal interfaces between monolayer transition-metal dichalcogenide semiconductors (such as WSe$_2$ or MoS$_2$) and magnetic substrates (such as EuO or CrI$_3$). Theoretical studies along these lines \cite{Qi:2015, Schwingenschlogl:2016, Scharf:PRL:2017, Zollner:2019, Zhang:2019, Xie:2019} have been validated by experiments demonstrating, e.g., enhanced valley splitting of WSe$_2$ and WS$_2$ monolayers on ferromagnetic EuS \cite{Zhao:NatNano:2017, Norden:NatComm:2019}, and zero-field valley splitting of MoSe$_2$ monolayers on ferromagnetic CrBr$_3$ \cite{Ciorciaro:PRL:2020}. In parallel, MPIs have also been shown to manifest as spin-dependent charge transfer and concomitant polarized photoluminescence in hybrid devices based on both CrI$_3$ \cite{Zhong:SciAdv:2017, Zhong:NatNano:2020} and CrBr$_3$ \cite{Lyons:NatComm:2020}.  In studies that measured the overall valley splitting of the A-exciton optical transition, the strength of the MPI could be characterized by an \textit{effective} magnetic field $B_{\rm eff}$, typically of order 10~T, acting on the semiconductor monolayer \cite{Zhao:NatNano:2017, Ciorciaro:PRL:2020, Zhong:SciAdv:2017}.  

While the notion of an overall effective field $B_{\rm eff}$ is certainly convenient for characterizing MPIs, it  obscures the very real possibility that the effects on spin-up and spin-down bands in the nonmagnetic monolayer (or equivalently, effects in the $K$ and $K'$ valleys) could be significantly different in magnitude.  Indeed, given that MPIs originate in the spin-dependent coupling between the underlying electronic band structures of the proximal materials, and that ferromagnets typically possess spin-polarized band structures, there is no \textit{a priori} reason to expect that the influence on the $K$ and $K'$ valleys should be equal and opposite, as it is for the case of real applied magnetic fields. 

To address this important question, we perform magnetic circular dichroism (MCD) and polarization-resolved reflection spectroscopy of van der Waals heterostructures comprising a MoSe$_{2}$ monolayer and few-layer CrBr$_{3}$. Pronounced valley splittings at zero applied magnetic field are observed at both the A- and B-exciton resonances of the MoSe$_2$ monolayer, which arise from MPIs with the ferromagnetic CrBr$_3$.  Square magnetic hysteresis loops of the induced MCD and spatially-resolved MCD images indicate that the CrBr$_3$ is a single magnetic domain. However, valley-resolved spectroscopy reveals that MPIs in this system are indeed markedly \textit{asymmetric}, and therefore act very differently from that of an effective magnetic field -- namely, the exciton shifts in the $K$ and $K'$ valleys of the MoSe$_2$ are very different in magnitude. Density-functional calculations support these findings, and indicate that strongly asymmetric MPIs originate from the spin-dependent hybridization of closely overlapping bands in the proximal layers, and are likely a general feature of such hybrid systems.

\begin{figure*}[t!]
\centering
\includegraphics[width=0.95 \textwidth]{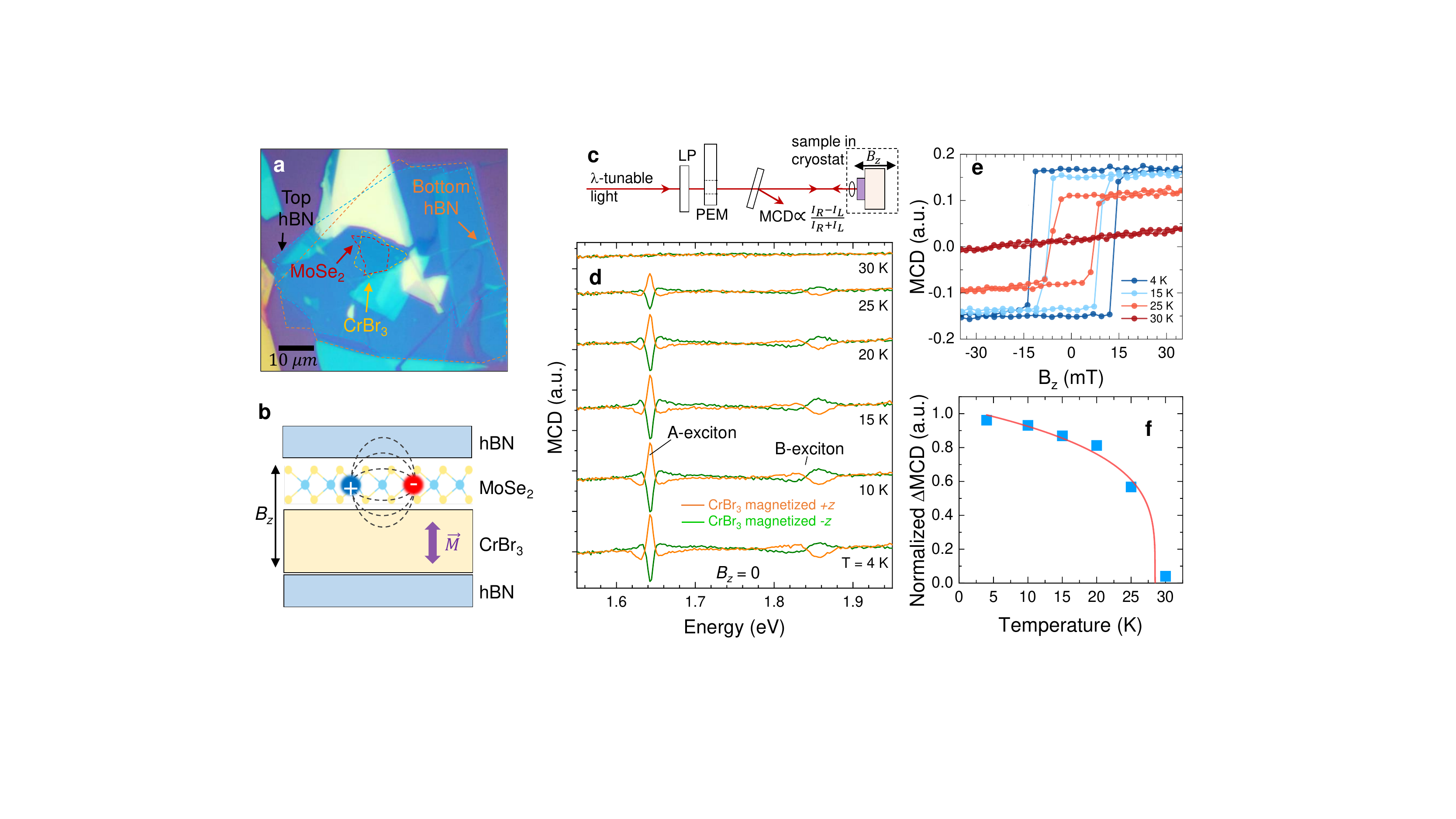}
\caption{\label{fig1} \textbf{Strong magnetic proximity interactions in a MoSe$_{2}$/CrBr$_{3}$ van der Waals heterostructure.}  \textbf{a}, Optical image of an hBN-encapsulated heterostructure.  \textbf{b}, Layer schematic. The MoSe$_2$ is a single monolayer, the CrBr$_3$ is 5 layers thick, and the hBN slabs are $\approx$10~nm thick. \textbf{c}, Schematic of the magnetic circular dichroism (MCD) experiment. Wavelength-tunable light, from either a filtered xenon lamp or a Ti:sapphire laser, was linearly polarized (LP), modulated between right- and left-circular polarization by a photoelastic modulator (PEM), and reflected from the sample. The normalized intensity difference, MCD=$(I_R - I_L)/(I_R + I_L)$, was measured via lock-in techniques.   \textbf{d}, MCD spectra measured at zero applied magnetic field ($B_z=0$), at temperatures above and below the ferromagnetic Curie temperature of CrBr$_3$ ($T_C \approx $28~K; curves offset for clarity). Orange and green traces were acquired after ramping $B_z$ to zero from large positive and negative values, respectively. When $T<T_C$, this permanently magnetizes the CrBr$_3$ in the $+\hat{z}$ or $-\hat{z}$ direction, respectively. Magnetic proximity interactions are revealed by the large MCD signals appearing at both the A- and B-exciton resonances of the MoSe$_2$ monolayer ($\approx$1.64 eV and 1.85 eV), which appear only below $T_C$.  \textbf{e}, Magnetic hysteresis of the MCD signal, acquired at the A-exciton resonance (1.642 eV), at different temperatures.  \textbf{f}, Temperature-dependent MCD signal at $B_{z} =0$.  The red line shows a fit to $(1-T/T_{C})^\beta$, using $T_C$=28~K and $\beta$=0.25.}
\end{figure*}

Figures 1a and 1b show an example of a MoSe$_2$/CrBr$_3$ heterostructure used in this study. All individual layers were mechanically exfoliated and stacked within an argon glovebox using standard dry transfer techniques (see Methods). The MoSe$_{2}$ monolayer was placed on few-layer CrBr$_{3}$ and sandwiched between hexagonal boron nitride (hBN) slabs to maintain high optical quality. CrBr$_{3}$ is a 2D van der Waals magnet exhibiting out-of-plane ferromagnetic order below its Curie temperature \cite{Chen:Science:2019, Kim:PNAS:2019, Soriano:2020, Mak:2019}. The structures were assembled on Si substrates, mounted on a confocal microscope probe, and loaded into the variable-temperature helium insert of a magneto-optical cryostat. Magnetic fields $B_z$ could be applied normal to the sample plane in the Faraday geometry. MCD spectroscopy (depicted in Fig. 1c, and described in Methods) was used to characterize the overall splitting between $K$ and $K'$ valleys in the MoSe$_2$ monolayer, due to MPIs with the CrBr$_3$ layer. To \textit{separately} resolve the energy shifts induced at the $K$ and $K'$ valleys of the MoSe$_2$, the spectra of right- and left-circularly polarized reflected light were individually measured. 

\begin{figure*}[t!]
\centering
\includegraphics[width=0.90\textwidth]{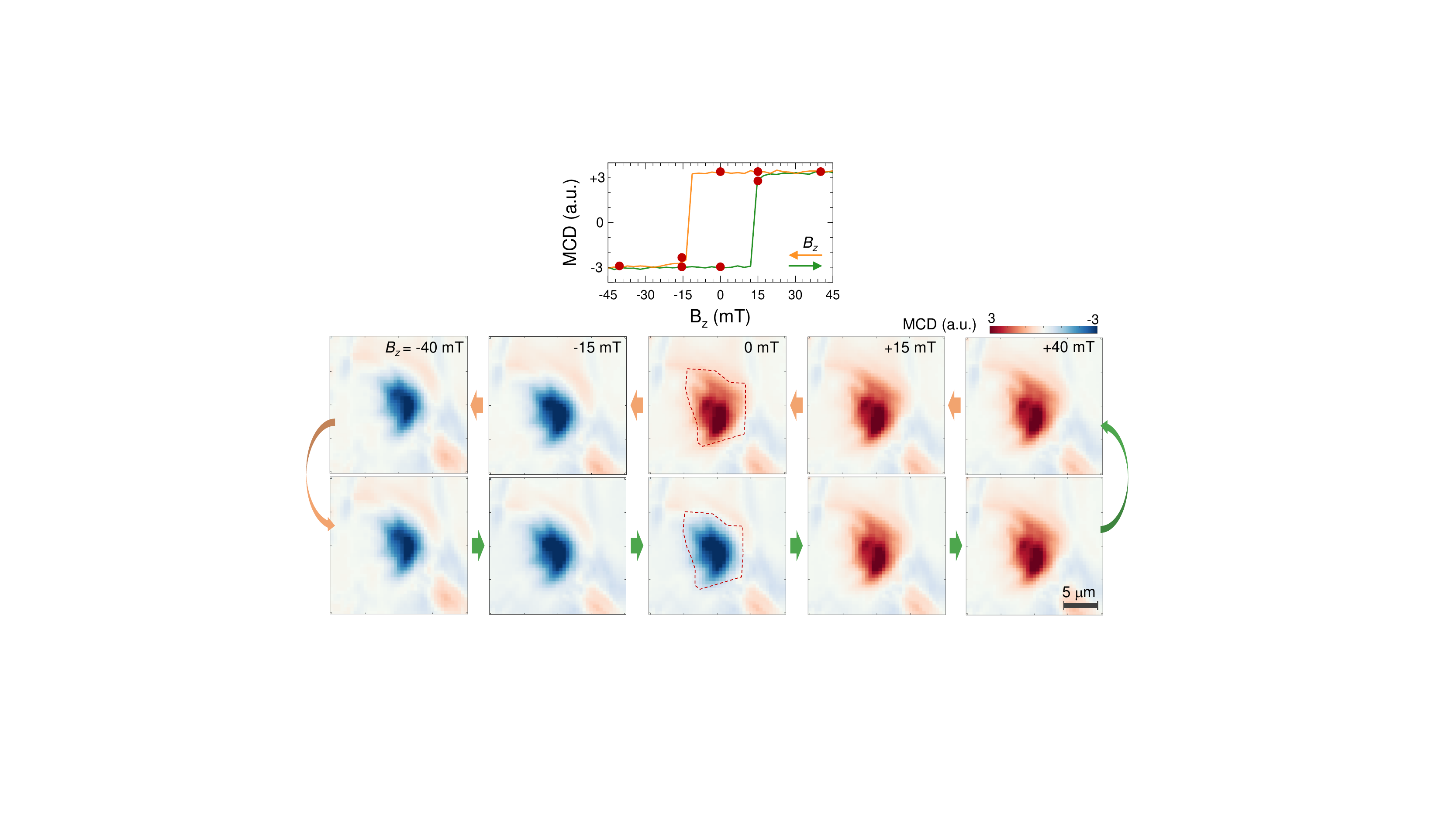}
\caption{\label{fig2} \textbf{Spatially imaging MPIs in a MoSe$_2$/CrBr$_3$ heterostructure.} The ten panels show MCD images, acquired at 4~K using 1.642 eV probe light (i.e., at the A-exciton of MoSe$_2$) at different applied fields $B_z$ around a typical magnetic hysteresis loop. The red dashed line shows the MoSe$_{2}$/CrBr$_{3}$ heterostructure region. The square hysteresis loop and MCD images indicate that the CrBr$_3$ behaves as a single magnetic domain. MCD images from a different MoSe$_2$/CrBr$_3$ structure are shown in Extended Data Fig. 2.}
\end{figure*}

The presence of strong MPIs in these structures is revealed by the temperature-dependent MCD spectra shown in Fig. 1d. MCD, which detects the intensity difference between right- and left-circularly polarized light (in reflection or transmission), is inherently sensitive to phenomena that break time-reversal symmetry, such as magnetization. Crucially, all spectra in Fig. 1d were acquired at $B_z=0$, after magnetizing the ferromagnetic CrBr$_3$ along the $+\hat{z}$ or $-\hat{z}$ direction. Below the Curie temperature ($T_C \approx 28$~K), pronounced MCD signals remain at $B_z$=0, at both of the fundamental A- and B-exciton resonances of the MoSe$_2$ monolayer, indicating the presence of MPIs.  Flipping the CrBr$_3$ magnetization from $+\hat{z}$ to $-\hat{z}$ inverts the MCD signals (as expected), and the zero-field MCD signals disappear above $T_C$. The particular lineshapes of the MCD resonances follow the derivatives of the A- and B-exciton resonances as measured in reflection (shown below in Fig. 3).

Figure 1e shows hysteresis loops of the MCD signal, acquired at the peak of the A-exciton resonance, when $B_z$ is varied between $\pm$40~mT, further confirming MPIs in these heterostructures. These data were taken using a wavelength-filtered xenon lamp, and the focused spot was approximately the same size as the sample itself ($\approx 4 \times 8 ~\mu$m). As such, the square hysteresis loops observed at low temperatures indicates that the entire CrBr$_3$ flake behaves as a single magnetic domain that switches uniformly, and is devoid of multi-domain switching phenomena. The hysteresis loops collapse with increasing temperature and disappear above $T_C$, in line with expectations. The MCD signal versus temperature, shown in Fig. 1f, is characteristic of the magnetization of ferromagnetic materials, and can be fit to the functional form $(1-T/T_C)^\beta$, using $T_C$=28~K and $\beta = 0.25$ (red line), consistent with expectations of a Heisenberg spin system \cite{Fisher:critical:1974, Gilbertini:2019}. A set of MCD hysteresis loops acquired using probe light tuned to the B-exciton resonance shows essentially similar behavior, as expected (see Extended Data Fig. 1).

Spatially-resolved images of the MCD over the entire MoSe$_2$/CrBr$_3$ heterostructure (see Fig. 2) confirm that MPIs are approximately uniform over the entire structure, and that the CrBr$_3$ layer behaves effectively as a single-domain magnet. For imaging, the probe light was derived from a continuous-wave Ti:S laser, and $\sim$1 $\mu$m spatial resolution is achieved. MCD images at various applied $B_z$ around the hysteresis loop show a single domain structure, even near the coercive fields ($B_z \simeq \pm 15$~mT).  MCD images of other vdW heterostructures having thicker CrBr$_3$ showed multi-domain magnetic structure near the switching fields, as shown in Extended Data Fig. 2.

\begin{figure*}[t!]
\centering
\includegraphics[width=0.90\textwidth]{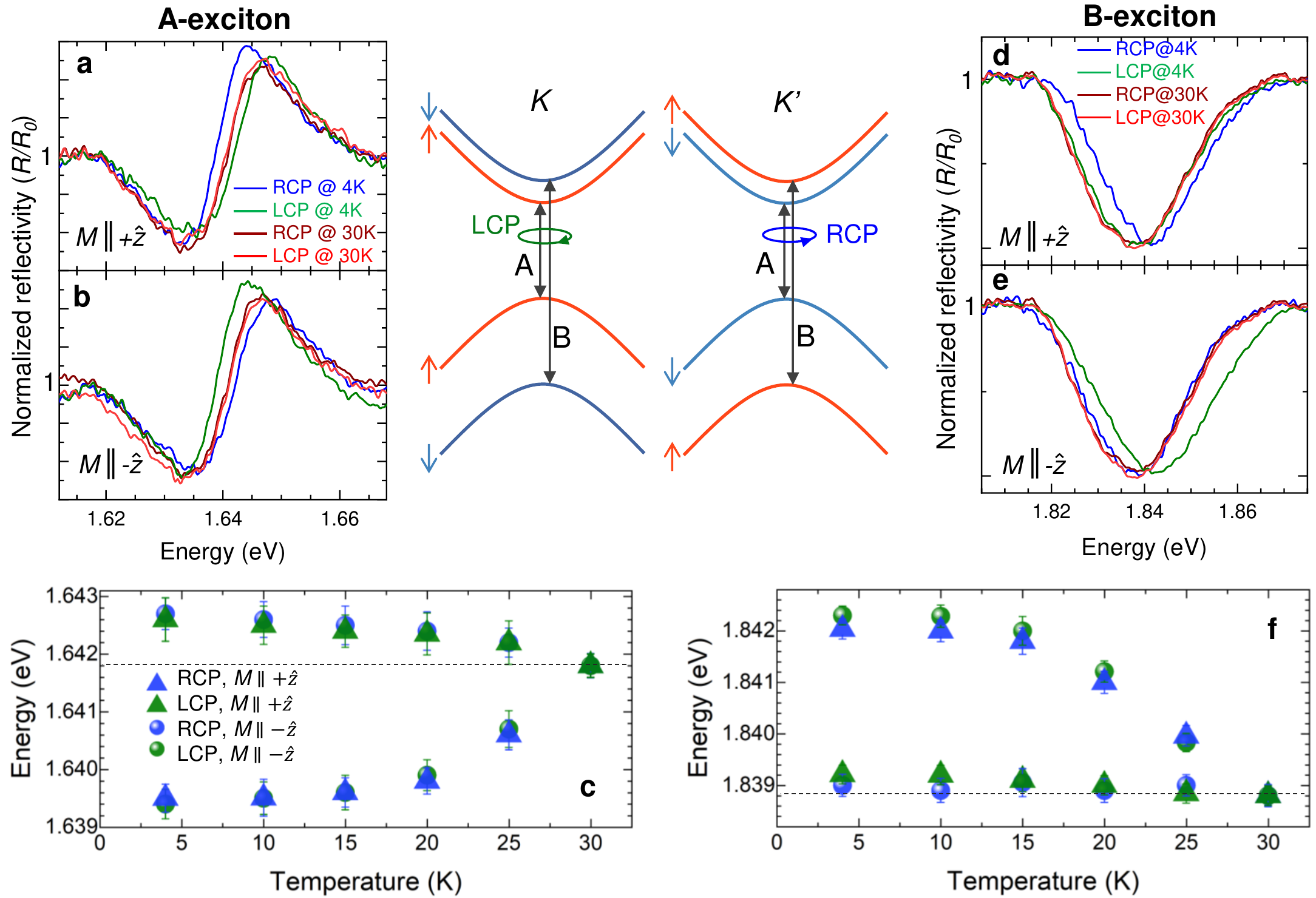}
\caption{\label{fig3} \textbf{Asymmetric MPIs in MoSe$_2$/CrBr$_3$ heterostructures.} \textbf{a}, Comparing the MPI-induced shift of the MoSe$_2$ A-exciton in the $K$ and $K'$ valleys at low temperature (4K), as measured by the reflection spectra of LCP and RCP light (green and blue traces, respectively). The CrBr$_3$ is magnetized along $+\hat{z}$. Red and orange traces show reference spectra acquired at 30~K (above $T_C$), where the CrBr$_3$ is unmagnetized. The MPI-induced shift of the A-exciton in the $K'$ valley (RCP light) is approximately twice as large as that in the $K$ valley (LCP light). All spectra were acquired at $B_z=0$. \textbf{b}, Same, but with the CrBr$_3$ magnetized along $-\hat{z}$; now the MPI-induced shift in the $K$ valley is larger. \textbf{c}, The temperature dependence of the MPI-induced valley shifts of the A-exciton, for both circular polarizations and for both $\pm \hat{z}$ CrBr$_3$ magnetization. \textbf{d-f}, A similar comparison of the MPI-induced energy shifts at the B-exciton resonance of MoSe$_2$. At both A- and B-exciton optical resonances, the MPI-induced shifts exhibit strong asymmetry, in contrast to the effects of a real applied magnetic field.}
\end{figure*}

Having established the presence of strong MPIs in these hybrid structures, we now turn to the main result:  In stark contrast to the influence of real magnetic fields, which shift time-reversed pairs of bands in the $K$ and $K'$ valleys of MoSe$_2$ equally and in opposite directions, we find that MPIs in these heterostructures are strongly asymmetric. That is, proximity-induced shifts of the exciton resonances have markedly different magnitudes in the $K$ and $K'$ valleys. To separately analyze these shifts, we use circular-polarization-resolved, and therefore valley-resolved, optical reflection spectroscopy. (Note that MCD is less sensitive to such asymmetry, since it detects the \textit{difference} between right- and left-circular polarization, and not the shift in each valley separately). 

Figure 3a demonstrates this asymmetry. The blue and green traces show the 4~K reflection resonance of the A-exciton in the MoSe$_2$ monolayer, for RCP and LCP light ($K'$ and $K$ valleys) respectively, when the CrBr$_3$ is magnetized along the $+\hat{z}$ direction. To serve as a common reference and control experiment, the red and orange traces were acquired at 30~K (above $T_C$), where the CrBr$_3$ is unmagnetized (and indeed, the red and orange spectra are identical to within experimental noise). Clearly, the magnitude of the energy shift at low temperature is noticeably larger in $K'$ (blue trace) than in $K$ (green trace). This behavior contrasts with the valley Zeeman shifts arising from real applied magnetic fields, which shift the bands equally and in opposite directions \cite{Stier:2016}. Importantly, Fig. 3b shows that the LCP/RCP ($K/K'$) asymmetry inverts when the CrBr$_3$ is magnetized along the opposite direction ($-\hat{z}$), confirming that the asymmetry is due to MPI interactions. 

The individual energy shifts in the $K$ and $K'$ valleys can be determined from fits of the reflection resonances to a complex-Lorentzian lineshape (see Methods), and tracked with good precision. The evolution of the asymmetric energy splitting as a function of temperature is shown in Fig. 3c.  The asymmetry exceeds a factor of two, with an energy redshift of $\approx$2~meV in the $K'$ valley that is accompanied by an opposite blueshift of only $\approx$1~meV in the $K$ valley (for the case of $+\hat{z}$ CrBr$_3$ magnetization).  The \textit{total} energy splitting of $\sim$3~meV that is observed at low temperatures is commensurate with the (symmetric) valley splitting expected from a magnetic field of approximately 13~T, assuming a neutral A-exciton g-factor $g_{ex}= 4$, in very good correspondence with prior work by Ciorciaro \textit{et al.} \cite{Ciorciaro:PRL:2020}. We emphasize that the asymmetric shifts are not an artifact of any temperature-dependent band-gap shift arising from the 30~K `control' spectra; redshifts of the bandgap between 4~K and 30~K are negligible, and furthermore if present would only make the observed asymmetry even more pronounced.

Interestingly, a rather different asymmetry is observed at the B-exciton of MoSe$_2$, as shown in Fig. 3d. Here, MPIs induce a marked energy shift in the $K'$ valley (when the CrBr$_3$ is magnetized along $+\hat{z}$), but almost no discernible shift in the $K$ valley. Taken together, we therefore conclude that characterizing the valley splitting by a simple effective magnetic field $B_{\rm eff}$ is of limited utility, as it does not capture the marked asymmetry that clearly arises from MPIs. Figure 3e confirms that the $K/K'$ asymmetry  at the B-exciton inverts when the CrBr$_3$ magnetization is flipped, and Fig. 3f shows the evolution of the MPI-induced energy shifts of the B-exciton as a function of temperature.

\begin{figure*}[t!]
\centering
\includegraphics[width=0.99\textwidth]{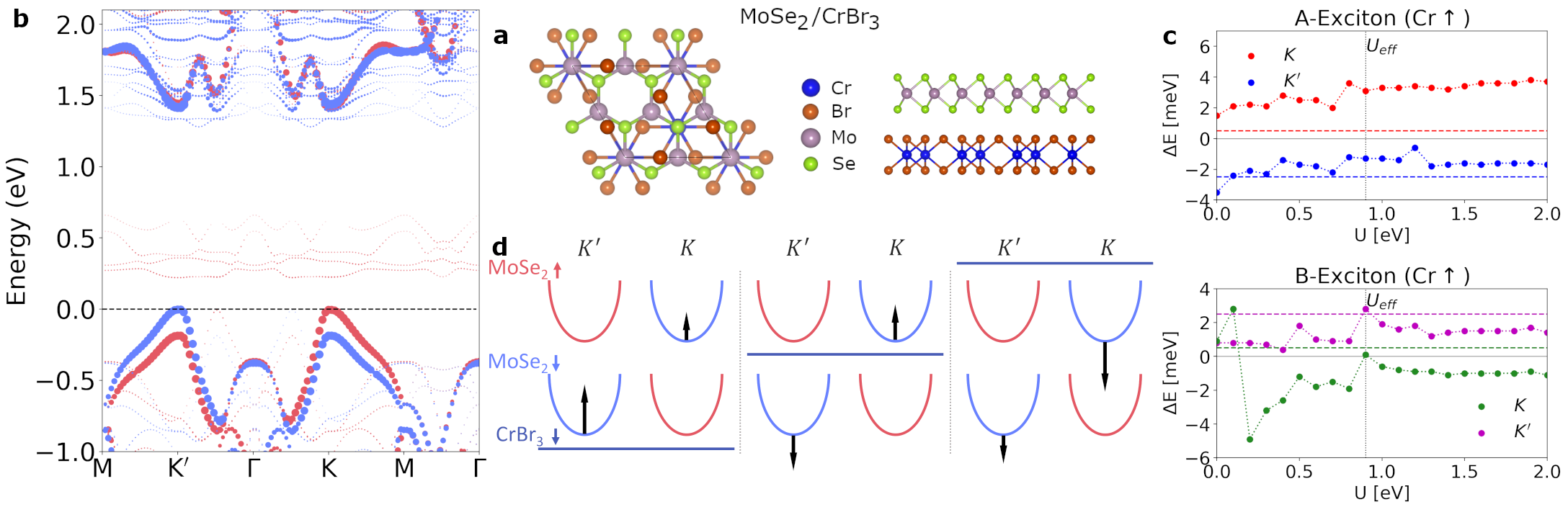}
\caption{\label{fig4}
\textbf{Crystal structure and theoretical electronic band dispersion of MoSe$_2$/CrBr$_3$, showing asymmetric valley shifts due to MPI.} \textbf{a}, The crystal structure of the MoSe$_2$/CrBr$_3$ bilayer as viewed from above, and along the in-plane $b$-axis. Here, the two monolayers are stacked such that the magnetic chromium sites sit directly below a molybdenum or a selenium atom of MoSe$_2$ (``AA stacking'') . \textbf{b}, The unfolded electronic band structure of the MoSe$_{2}$/CrBr$_{3}$ heterostructure, with CrBr$_3$ in the ferromagnetic phase and magnetized along $+\hat{z}$.  The size of the red (blue) dots is proportional to the fractional weights of the spin-up (-down) MoSe$_2$-layer projection in the unfolded Brillouin zone. \textbf{c}, Calculated energy shifts of the A- and B-exciton transition energies in the ferromagnetic phase of CrBr$_3$ relative to a planar-antiferromagnetic state, over a wide range of Hubbard $U_{eff}$ (adjusting $U_{eff}$ shifts the CrBr$_3$ bands with respect to the MoSe$_2$ bands). The planar-antiferromagnetic reference phase is used as an analogue to the experimental high-temperature non-magnetic phase of CrBr$_3$. Valley-asymmetric shifts due to MPI are nearly universal, with clear overall trends as a function of $U_{eff}$.  Calculations on other stacking arrangements are shown in the Extended Data Figs. 4-7. \textbf{d}, Diagrams illustrating the origin of the asymmetric valley shifts due to MPI: The polarized $t_{2g}$ bands of CrBr$_3$ (represented here by a single level), which are very close in energy to the MoSe$_2$ conduction bands, primarily repel  the like-spin conduction bands of MoSe$_2$, leading to $K/K'$ asymmetry.  The magnitude and sign of the shifts, indicated by black arrows, depends sensitively on the precise alignment of the levels. }
\end{figure*}

Comparing the energy shifts of the A- and B-excitons in Figs. 3c and 3f reveals another unexpected aspect of asymmetric MPIs: For a given CrBr$_3$ magnetization, the dominant energy shifts for both A- and B-exciton transitions are not only of different sign, but are observed in the same valley. Considering the spin-dependent level structure of the conduction and valence bands in MoSe$_2$ monolayers (see band diagram in Fig. 3), it might be anticipated that a spin-specific proximity interaction affecting (for example) only the spin-up bands, should cause shifts in the $K$ valley of the A-exciton, and in the $K'$ valley of the B-exciton. These data demonstrate that the coupling between the electronic band structures of the two constituent materials is more complex. 

To provide insight into the asymmetric valley shifts driven by MPIs, we examined the electronic band structure of monolayer MoSe$_2$ stacked on a single layer of CrBr$_3$ (Fig. 4a) within the framework of density functional theory (DFT; see Methods). To most directly compare the MPI-induced energy shifts with experimental data (which includes control measurements in the non-magnetic phase of CrBr$_3$), we considered two cases: (i) CrBr$_3$ with out-of-plane ferromagnetic order, to capture the low-temperature phase, and (ii) CrBr$_3$ with in-plane antiferromagnetic spin configuration, to simulate the high-temperature non-magnetic state and serve as a reference. We note that since the band energies are sensitive to the overall electronic environment and band alignment between the CrBr$_3$ and MoSe$_2$ layers (see for example Fig. 4b, or Extended Data Fig. 3), one cannot simply compare the band shifts with and without the CrBr$_3$ layer, due to intrinsic interlayer coupling even in the non-magnetic phase. 

Figure 4c shows the calculated shifts of the optical transition energies in the $K$ and $K^{\prime}$ valleys due to MPIs as function of band alignment between MoSe$_2$ and CrBr$_3$ layers (as controlled by the Hubbard $U_{eff}$; see Methods), with the experimental values overlaid. A significant valley asymmetry is clearly revealed at both of the fundamental A and B optical transitions, in line with our valley-resolved reflection spectra. Specifically, a $U_{eff}$ value of 0.9 eV simultaneously yields the experimental band gap of CrBr$_3$ and the measured shifts of the optical transition energies. Here, our DFT results indicate asymmetric shifts with the same sign structure and valley dependence for both CrBr$_3$ magnetization directions. In addition, the magnitudes of the calculated shifts are in reasonable and qualitative accord with experiment. Further refinements of the DFT values, beyond the scope of this work, may be obtained by employing GW quasiparticle corrections and calculating the full exciton spectrum within the Bethe-Salpeter equation framework. However, we stress that these calculations even at the DFT level already capture the essential trends and marked asymmetries shown in the experimental data. 

The asymmetric valley splittings due to MPIs can be understood as follows. Due to the band alignment between the CrBr$_3$ and MoSe$_2$ layers, the conduction states of both layers directly overlap and hybridize, as indicated by the weaker band features in the MoSe$_2$ projected dispersions shown in Fig. 4b. Extended Data Fig. 3 explicitly shows the electronic bands projections for both CrBr$_3$ and MoSe$_2$ layers revealing the significant mixing between the electronic states of the layers. This hybridization drives resonant avoided crossing phenomena between same-spin bands, thereby producing strong band shifts, with magnitude and sign sensitively dependent on the detailed alignment of the constituent levels. Figure 4d shows a simplified illustration that highlights the essential physics: For various band alignments, like-spin MoSe$_2$ conduction bands shift to avoid the upper (spin-polarized) $t_{2g}$ bands of ferromagnetic CrBr$_3$, yielding valley-dependent shifts of the optical transitions that differ qualitatively from a picture of simple Zeeman splitting. Due to the avoided-crossing behavior, the calculated band shifts have magnitude and sign that depend sensitively on the relative energies of the upper CrBr$_3$ bands and the MoSe$_2$ conduction bands, which marks a distinct departure from the effective magnetic field characterization typically employed to quantify MPIs. This behavior is directly seen in the calculated shifts (Fig. 4c) as peaks and troughs in the data, marking the energies at which the CrBr$_3$ states cross the MoSe$_2$ bands. (Note that the MoSe$_2$ \textit{valence} bands at $K$ and $K^{\prime}$ display an approximately symmetric valley splitting, as expected since they do not closely overlap with any states of the CrBr$_3$ layer). Extended Data Figs. 4-7 demonstrate how, within our DFT approach, the calculated energy shifts and the degree of asymmetry evolve for different stacking arrangements of the constituent CrBr$_3$ and MoSe$_2$ layers, for various band alignments. These results also suggest that the band alignment and inter-layer coupling of a given magnetic heterostructure may be quantified by the experimentally-measured sign and magnitude of asymmetric valley shifts. The extent to which twist angle between the MoSe$_2$ monolayer and the CrBr$_3$ affects the valley shifts in such hybrid structures has not been explored here, and remains an open question. 

By revealing the marked spin/valley asymmetries that can arise from MPIs in hybrid vdW structures, this work suggests routes toward selective control over \textit{specific} spin and/or valley states in monolayer TMD semiconductors via rational design of the component materials, and also their stacking arrangement. Together with the additional tunability of MPIs that can arise from electrostatic gating (as demonstrated in graphene-based hybrid vdW heterostructures \cite{Kawakami:2018}), and recent predictions of gate-tunable band topology in monolayer semiconductors \cite{Tong:2020}, these results complement the set of available tools with which future spin- and valley-dependent (opto)electronic devices may be engineered.  Moreover, our experimental data and DFT calculations strongly indicate that asymmetric magnetic proximity interactions are likely a universal aspect of hybrid vdW heterostructures, manifesting especially in material combinations where electronic bands overlap closely in energy. \\

\textbf{METHODS} \\

\textbf{Sample preparation.}  Monolayer MoSe$_{2}$ (2D Semiconductors) and hBN flakes were mechanically exfoliated from bulk crystals onto Si substrates under ambient conditions and moved into an argon glovebox.  O$_{2}$ and H$_{2}$O concentrations in the glovebox were maintained below 0.1 ppm. Few-layer flakes of CrBr$_{3}$ (HQ graphene) were exfoliated onto Si substrates inside the glovebox. MoSe$_{2}$/CrBr$_{3}$ heterostructures were assembled by standard dry transfer techniques \cite{Kim:transfer:2016} using polycarbonate stamps, and were sandwiched between hBN slabs. Then the samples were moved out of the glovebox and washed in chloroform to remove polymer residues. The thickness of the CrBr$_{3}$ (typically 5 layers) was identified by atomic force microscopy after the polymer residue was removed. The samples were mounted on a confocal microscope probe and loaded into the variable-temperature helium-flow insert of a 7~T superconducting magnet with direct optical access.\\

\textbf{MCD spectroscopy and imaging.} Broadband MCD spectroscopy was performed in a reflection geometry using wavelength-tunable narrowband light derived from a xenon white light source filtered through a 300~mm spectrometer. The probe light was intensity-modulated by mechanical chopper, and then modulated between right- and left-circular polarizations by a linear polarizer and photoelastic modulator (PEM).  The light was focused to a small spot (approximately $4 \times 8~\mu$m) on the sample by an aspheric lens that was controlled by a piezo nanopositioner (attocube). The back-reflected light from the sample was collected by the same lens and directed by a beamsplitter to an avalanche photodiode detector (as depicted in Fig. 1c). The signal was demodulated by two lock-in amplifiers, referenced to the chopper and PEM frequencies (137 Hz and 50 kHz, respectively). MCD is given by the normalized difference between the right- and left-circularly polarized reflected intensities, $(I_R - I_L)/(I_R + I_L)$.  For the MCD imaging experiments shown in Fig. 2, the probe light was derived instead from a wavelength-tunable continuous-wave Ti:sapphire ring laser, so that much better spatial resolution could be achieved ($\sim$1~$\mu$m focused spot size). The incident probe laser beam was coupled to a 2D galvo mirror scanner for raster scanning and spatial imaging.\\

\textbf{Polarization- (valley-) resolved reflection spectroscopy.}  The polarization-resolved reflection measurements were performed using the same confocal microscope described above.  Here, broadband white light from a xenon lamp was coupled into a single-mode fiber, then collimated by an achromatic lens, and then circularly polarized using a linear polarizer and quarter-wave plate. The white light was then focused to a $\sim$1 $\mu$m spot with an aspheric lens. The reflected light from the sample was collected by a multi-mode fiber and detected by a cooled charge-coupled device (CCD) detector. As is typical for reflection spectroscopy of vdW heterostructures, the A- and B-exciton resonances exhibit a complex-Lorentzian lineshape \cite{xfermatrix} (\textit{i.e.}, having both dispersive and absorptive components to the lineshape). Once a fit was established at a given temperature, the temperature-dependent shift of the transition energy (see Fig. 3) was accurately tracked by fixing all parameters except for the resonance energy.  Alternatively, letting all fit parameters float yielded essentially identical results, because the lineshape remains unchanged. \\

\textbf{Density functional theory calculations.} Density functional theory (DFT) based first-principles electronic structure calculations were carried out by using the pseudopotential projector-augmented wave method \cite{Kresse1999} implemented in the Vienna ab initio simulation package (VASP) \cite{Kresse1996, Kresse1993}. We used an energy cutoff of $300$ eV for the plane-wave basis set. Exchange-correlation effects were treated using the Perdew-Burke-Ernzerhof (PBE) GGA density functional \cite{Perdew1996}. A $15\times 15\times 1$ $\Gamma$-centered k-point mesh was used to sample the Brillouin zone. Spin-orbit coupling effects were included self-consistently. An effective Hubbard $U_{eff}$ was added to the Cr-3$d$ orbitals \cite{Dudarev} to control the band alignment and correct the band gap of the magnetic layer. A $U_{eff}$ of 0.9 eV was found to yield the experimental band gap of CrBr$_3$ and the asymmetric optical transitions simultaneously. The heterobilayer was constructed from a $1\times 1$ unit cell of CrBr$_3$ and a $2\times 2$ super cell of MoSe$_2$. To not alter the electronic structure of the MoSe$_2$ layer, we strained CrBr$_3$ to make both layers commensurate \cite{Kormanyos2015}. Three distinct stacking configurations were considered, as shown in the Extended Data. To ensure negligible interaction between the periodic images of the bilayer film, a large enough vacuum of 19 \AA~ in the $z$-direction was used. Finally, all atomic sites in the unit cell were relaxed simultaneously using a conjugate gradient algorithm to minimize energy with an atomic force tolerance of $0.01$ eV/\AA~ and a total energy tolerance of $10^{-6}$ eV. The unfolded band structure was obtained using the PyProcar package \cite{pyprocar}. \\

\textbf{Acknowledgements} We gratefully acknowledge Igor \v{Z}uti\'{c} and Bernhard Urbaszek for helpful discussions. Experimental studies at the NHMFL were supported by the Los Alamos LDRD program. The NHMFL is supported by National Science Foundation (NSF) DMR-1644779, the State of Florida, and the U.S. Department of Energy (DOE). Computational studies were supported in part by the Center for Integrated Nanotechnologies, a DOE BES user facility, in partnership with the LANL Institutional Computing Program for computational resources. \\

\newpage

\setcounter{figure}{0} 

\begin{figure*}[t!]
\centering
\includegraphics[width=0.40\textwidth]{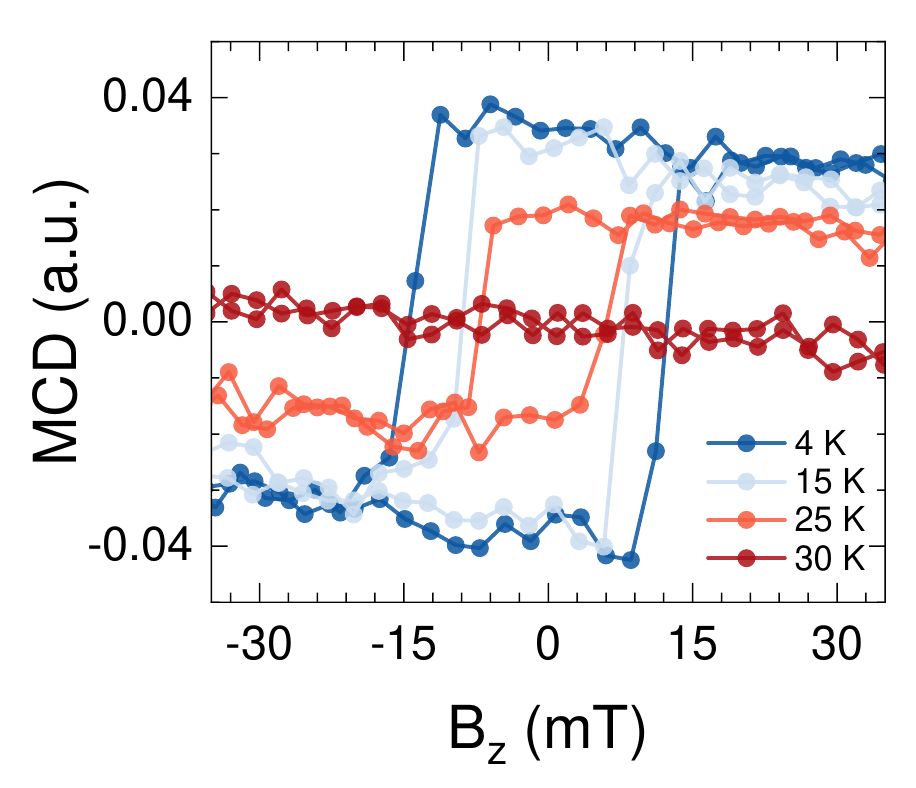}
\caption{\textbf{Extended Data Figure 1}.  Hysteresis loops of the MCD signal, acquired at the B-exciton transition of the MoSe$_2$ monolayer (photon energy = 1.83~eV; \textit{cf.} Fig. 1e of the main text).}
\end{figure*}

\newpage

\begin{figure*}[t!]
\centering
\includegraphics[width=0.60\textwidth]{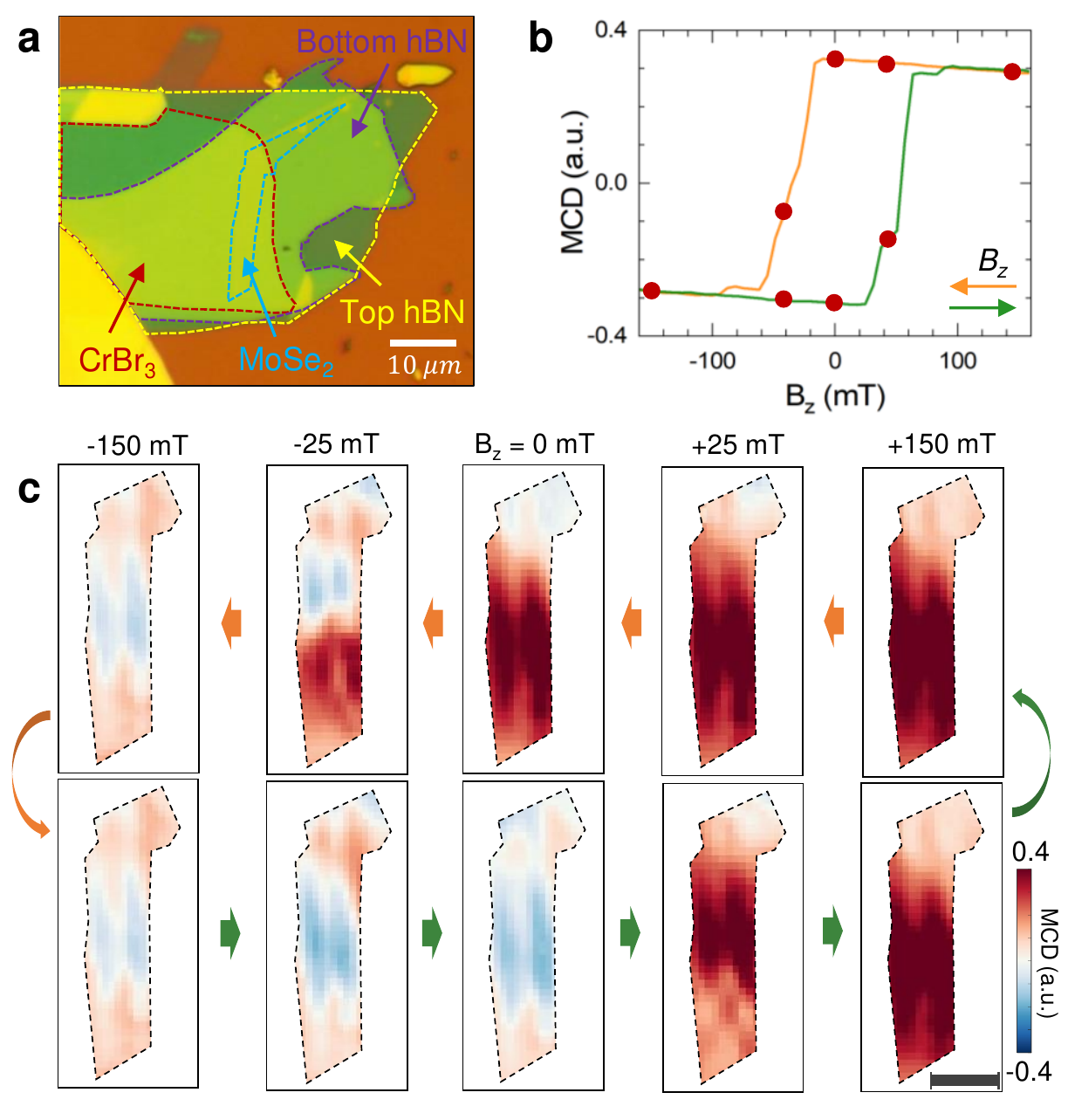}
\caption{\textbf{Extended Data Figure 2}. MCD images of magnetic proximity effects on a different MoSe$_2$/CrBr$_3$ heterostructure. a) Optical microscope image of the vdW structure.  b) Magnetic hysteresis of the MCD signal, using light tuned to the A-exciton transition of the MoSe$_2$ monolayer. c) MCD images acquired at 4~K, using probe light tuned to the peak of the A-exciton MCD resonance (photon energy = 1.64~eV), at selected applied magnetic fields $B_z$. The MoSe$_2$/CrBr$_3$ heterostructure region is indicated by the black dashed lines.}
\end{figure*}

\newpage

\begin{figure*}[t!]
\centering
\includegraphics[width=0.9\textwidth]{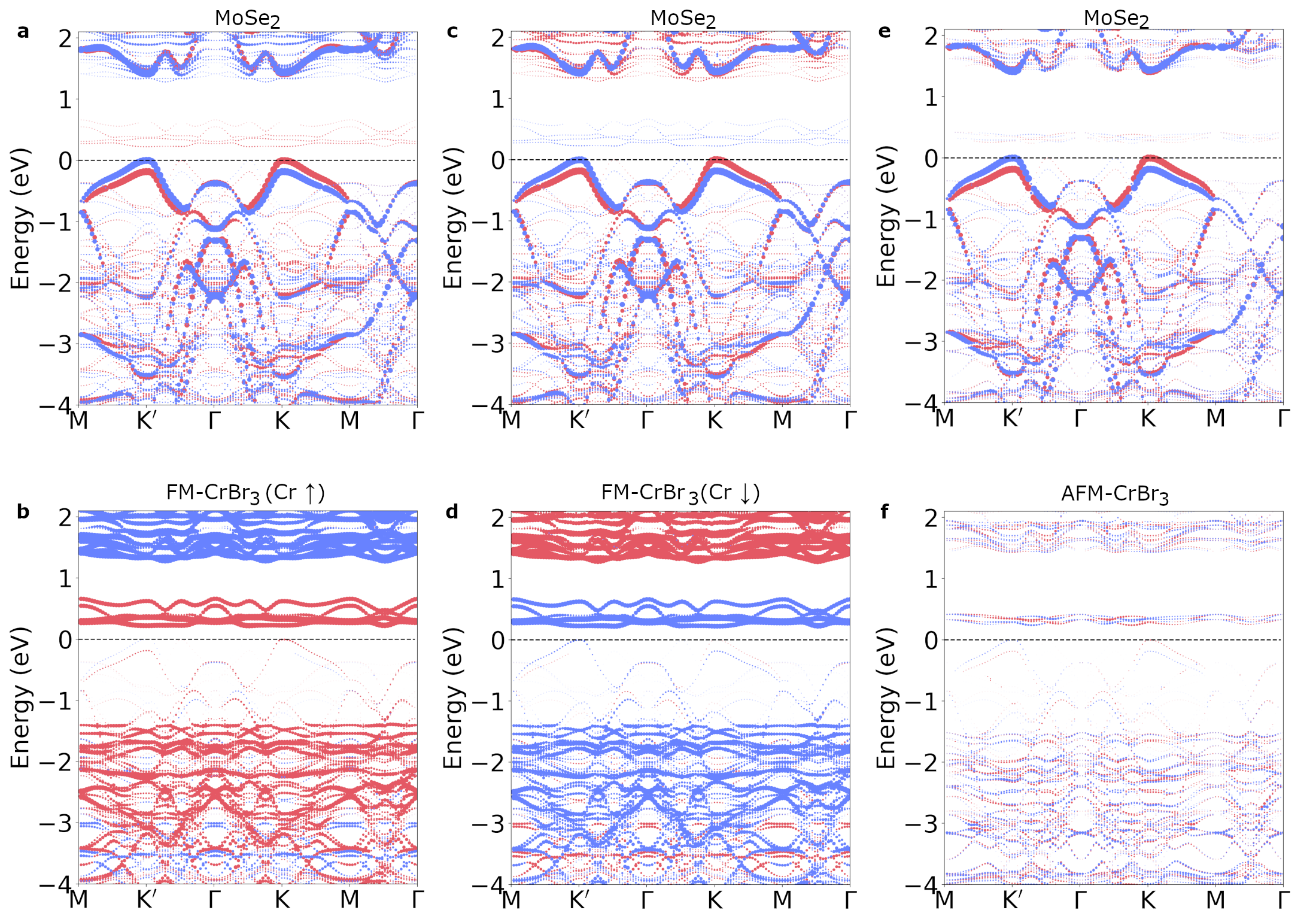}
\caption{\textbf{Extended Data Figure 3}. Theoretical electronic band structures for various CrBr$_3$ magnetic configurations. The unfolded electronic band structure of the MoSe$_2$/CrBr$_3$ heterostructure for CrBr$_3$ ferromagnetically polarized along $+\hat{z}$ (\textbf{a}-\textbf{b}), $-\hat{z}$ (\textbf{c}-\textbf{d}), and in the planar-antiferromagnetic phase (\textbf{e}-\textbf{f}). Top and bottom panels show the MoSe$_2$ and CrBr$_3$ layer projections, respectively. The sizes of the red (blue) dots are proportional to the fractional weights of the spin-up (-down) MoSe$_2$ and CrBr$_3$ layer projections, respectively. In the first two columns, the CrBr$_3$ conduction states clearly cut through the MoSe$_2$ unoccupied bands, thereby generating substantial level mixing and repulsion. The significant hybridization between these sets of bands is marked by the pronounced `shadow' of CrBr$_3$ bands in the MoSe$_2$ projected states (\textbf{a},\textbf{c}). In the right-most column (panels \textbf{e,f}), minimal band mixing between the layers is observed. Since the chromium magnetic moments lie in the plane, the electron overlap integrals connecting the CrBr$_3$ layer and the spin-polarized bands in MoSe$_2$ are significantly reduced. Furthermore, since there is no net magnetic moment in the planar-antiferromagnetic phase, the valley degeneracy is preserved.}
\end{figure*}

\newpage

\begin{figure*}[t!]
\centering
\includegraphics[width=0.9\textwidth]{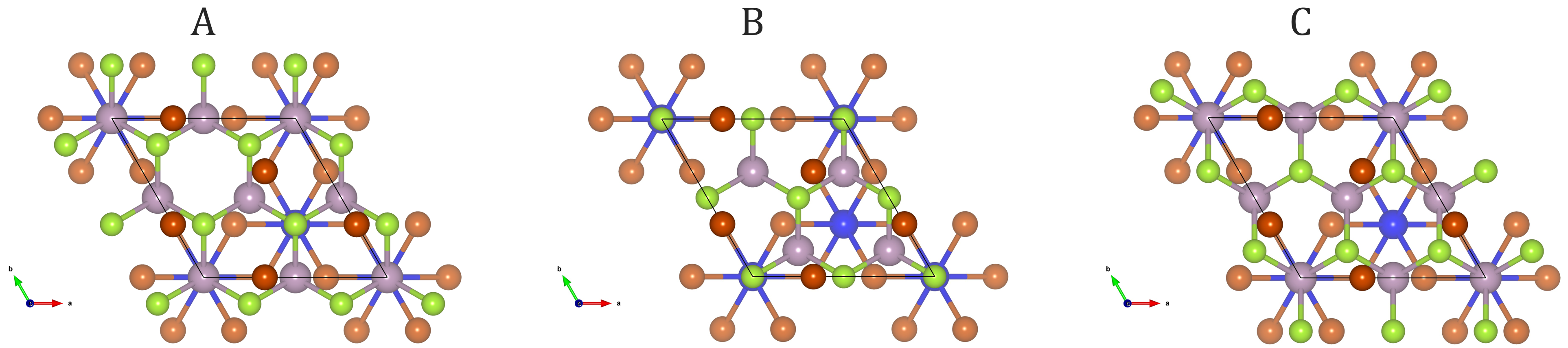}
\caption{\textbf{Extended Data Figure 4}. Crystal structure of MoSe$_2$/CrBr$_3$ for various stacking arrangements. The MoSe$_2$/CrBr$_3$ bilayer structure viewed from the top for AA, AB, and AC stacking configurations. The green, violet, orange, and blue spheres denote the selenium, molybdenum, bromine, and chromium, respectively. The black line indicates the unit cell boundary.}
\end{figure*}

\newpage

\begin{figure*}[t!]
\centering
\includegraphics[width=0.9\textwidth]{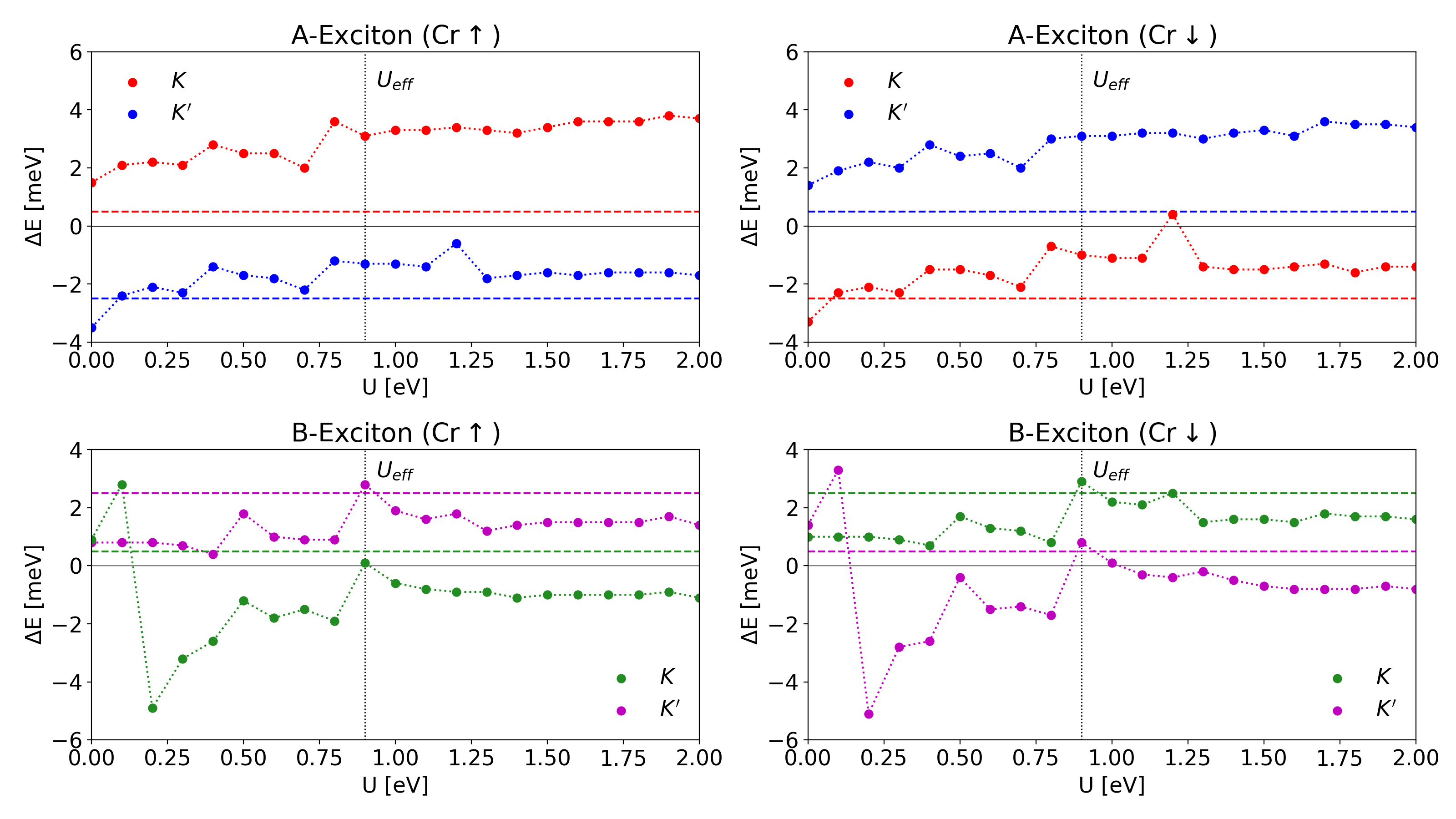}
\caption{\label{fig:AA} \textbf{Extended Data Figure 5}. Calculated shifts of the optical transition energies in the $K$ and $K^{\prime}$ valleys as a function of $U_{eff}$ for AA stacked MoSe$_2$/CrBr$_3$. As the effective Hubbard $U_{eff}$ on the chromium atomic sites is increased from 0 to 2 eV the optical transition energies in the $K$ and $K^{\prime}$ valleys (solid lines with dots) display a non-monotonic evolution for both $+\hat{z}$ and $-\hat{z}$ chromium spin polarizations. Since the relative band alignment between CrBr$_3$ and MoSe$_2$ states changes with increased $U_{eff}$, the resulting resonant avoided crossing phenomena produces a blue or red shift in the optical transition energies. A $U_{eff}=0.9$ eV is found to to simultaneously reproduce the experimental band gap of CrBr$3$ and qualitatively capture the asymmetry of the observed shifts of the measured optical transition energies (dashed lines).}
\end{figure*}

\newpage

\begin{figure*}[t!]
\centering
\includegraphics[width=0.9\textwidth]{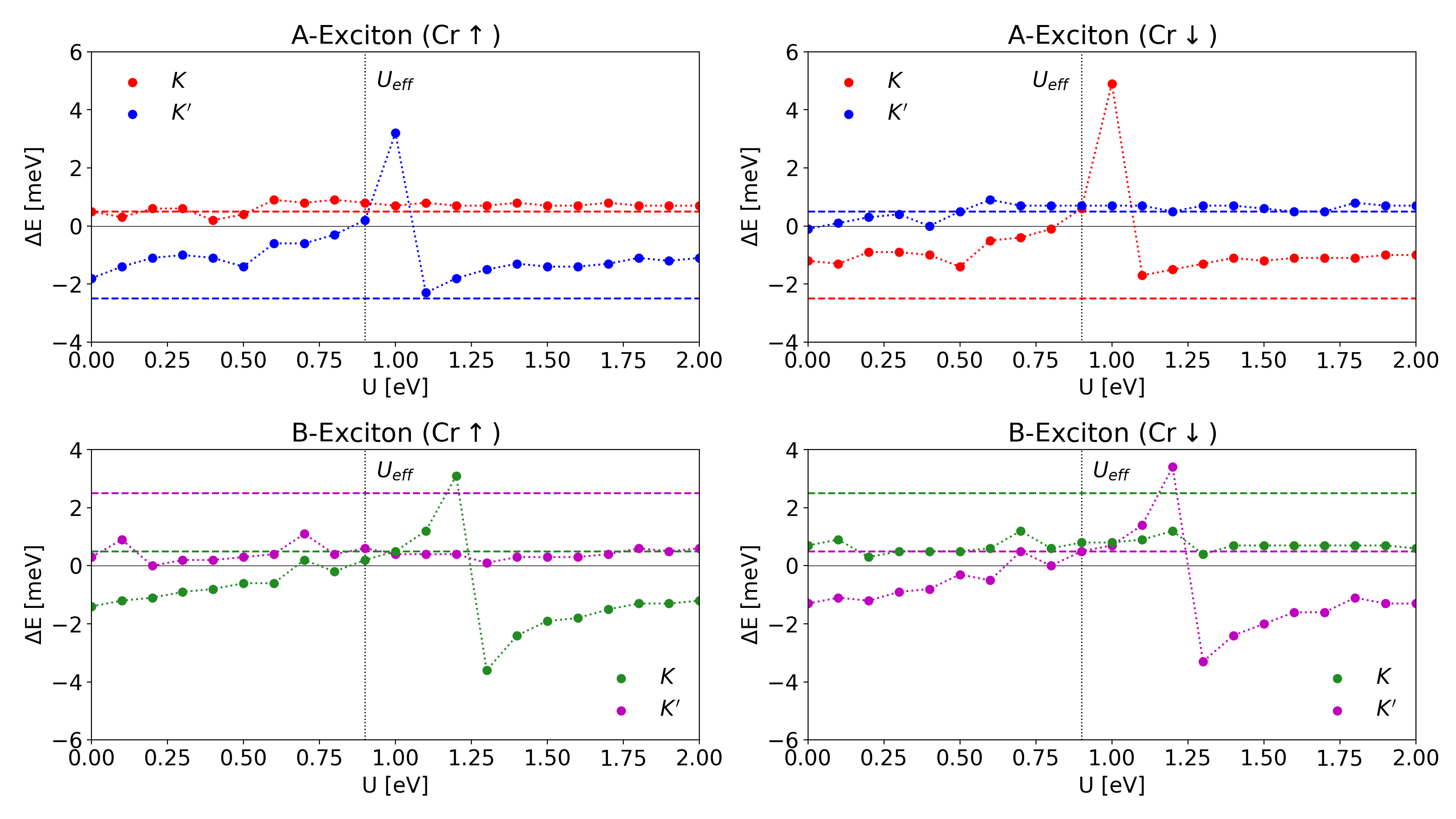}
\caption{\textbf{Extended Data Figure 6}. Same as Extended Data Fig.~\ref{fig:AA}, except for AB stacked MoSe$_2$/CrBr$_3$.}
\end{figure*}

\newpage

\begin{figure*}[t!]
\centering
\includegraphics[width=0.9\textwidth]{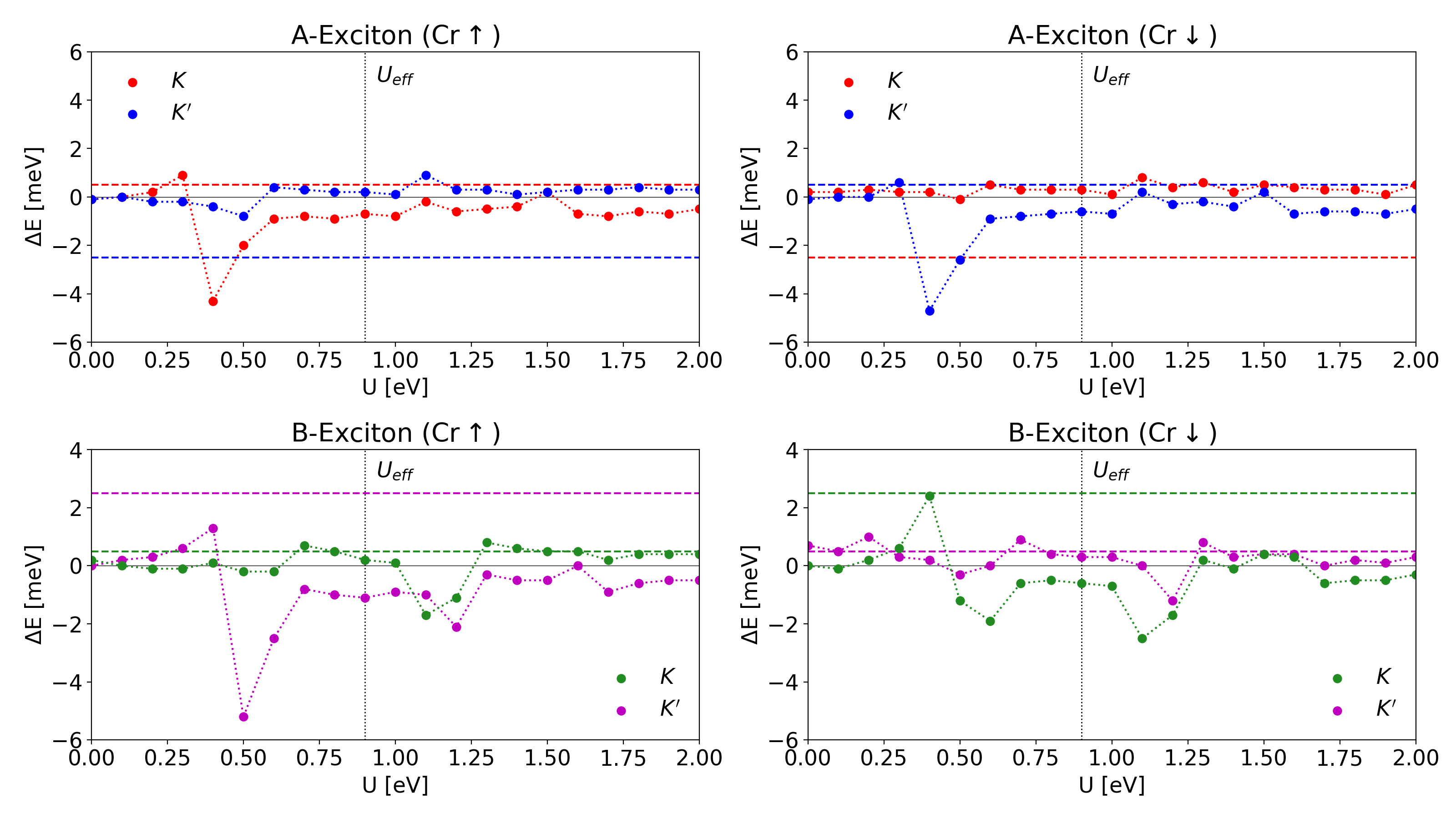}
\caption{\textbf{Extended Data Figure 7}. Same as Extended Data Fig.~\ref{fig:AA}, except for AC stacked MoSe$_2$/CrBr$_3$.}
\end{figure*}


\begin{thebibliography}{10}

\bibitem{Zutic:2019}\v{Z}uti\'{c}, I., Matos-Abiague, A., Scharf, B., Dery, H. \& Belashchenko, K. Proximitized Materials, \textit{Materials Today}, \textbf{22}, 85 (2019).

\bibitem{Gilbertini:2019} Gibertini, M., Koperski, M., Morpurgo, A. F. \& Novoselov, K. S. Magnetic 2D materials and heterostructures, \textit{Nat. Nanotechnol.} \textbf{14}, 408–419 (2019).

\bibitem{Mak:2019} Mak, K. F., Shan, J. \& Ralph, D. Probing and controlling magnetic states in 2D layered magnetic materials, \textit{Nat. Rev. Phys.} \textbf{1}, 646-661 (2019).

\bibitem{Zhao:NatNano:2017} Zhao, C., Norden, T., Zhang, P., Zhao, P., Cheng, Y., Sun, F., Parry, J. P., Taheri, P., Wang, J., Yang, Y., Scrace, T., Kang, K., Yang, S., Miao, G.-X., Sabirianov, R., Kioseoglou, G., Huang, W., Petrou, A. \& Zeng, H. Enhanced valley splitting in monolayer WSe$_{2}$ due to magnetic exchange field, \textit{Nat. Nanotechnol.} \textbf{12}, 757-762 (2017).

\bibitem{Zhong:SciAdv:2017} Zhong, D., Seyler Kyle, L., Linpeng, X., Cheng, R., Sivadas, N., Huang, B., Schmidgall, E., Taniguchi, T., Watanabe, K., McGuire Michael, A., Yao, W., Xiao, D., Fu K.-M. C. \& Xu, X. Van der Waals engineering of ferromagnetic semiconductor heterostructures for spin and valleytronics, \textit{Sci. Adv.} \textbf{3}, e1603113 (2017).

\bibitem{Norden:NatComm:2019} Norden, T., Zhao, C., Zhang, P., Sabirianov, R., Petrou, A. \& Zeng, H. Giant valley splitting in monolayer WS$_2$ by magnetic proximity effect, \textit{Nat. Commun.} \textbf{10}, 4163 (2019).

\bibitem{Ciorciaro:PRL:2020} Ciorciaro, L., Kroner, M., Watanabe, K., Taniguchi, T. \& Imamoglu, A. Observation of Magnetic Proximity Effect Using Resonant Optical Spectroscopy of an Electrically Tunable MoSe$_{2}$/CrBr$_{3}$ Heterostructure, \textit{Phys. Rev. Lett.} \textbf{124}, 197401 (2020).

\bibitem{Zhong:NatNano:2020} Zhong, D., Seyler, K. L., Linpeng, X., Wilson, N. P., Taniguchi, T., Watanabe, K., McGuire, M. A., Fu, K.-M. C., Xiao, D., Yao, W. \& Xu, X. Layer-resolved magnetic proximity effect in van der Waals heterostructures, \textit{Nat. Nanotechnol.} \textbf{15}, 187-191 (2020).

\bibitem{Xiao:PRL:2012} Xiao, D., Liu, G.-B., Feng, W., Xu, X. \& Yao, W. Coupled Spin and Valley Physics in Monolayers of MoS$_{2}$ and Other Group-VI Dichalcogenides, \textit{Phys. Rev. Lett.} \textbf{108}, 196802 (2012).

\bibitem{Xu:2014}Xu, X., Yao, W., Xiao, D. \& Heinz, T. F. Spin and pseudospins in layered transition metal dichalcogenides.  \textit{Nat. Phys.} \textbf{10}, 343-350 (2014).

\bibitem{Lazic:2016} Lazi\'{c}, P., Belashchenko, K. D. \& \v{Z}uti\'{c}, I. Effective gating and tunable magnetic proximity effects in two-dimensional heterostructures, \textit{Phys. Rev. B} \textbf{93}, 241401(R) (2016)

\bibitem{Kawakami:2018} Xu, J., Singh, S., Katoch, J., Wu, G., Zhu, T., \v{Z}uti\'{c}, I. \& Kawakami, R. Spin inversion in graphene spin valves by gate-tunable magnetic proximity effect at one-dimensional contacts, \textit{Nat. Commun.} \textbf{9}, 2869 (2018).

\bibitem{MakShan:2016} Mak, K. F. \& Shan, J. Photonics and optoelectronics of 2D semiconductor transition metal dichalcogenides, \textit{Nat. Photon.} \textbf{10}, 216-226 (2016).

\bibitem{Urbaszek:2018} Wang, G., Chernikov, A., Glazov, M. M., Heinz, T. F., Marie, X., Amand, T. \& Urbaszek, B. Excitons in atomically thin transition metal dichalcogenides, \textit{Rev. Mod. Phys.} , \textbf{90}, 021001 (2018).

\bibitem{Strano:2012} Wang, Q. H., Kalantar-Zadeh, K., Kis, A., Coleman, J. N. \& Strano, M. S. Electronics and optoelectronics of two-dimensional transition metal dichalcogenides, \textit{Nature Nanotechnology} 7, 699-712 (2012).

\bibitem{Schaibley:2016} Schaibley, J. R., Yu, H., Clark, G., Rivera, P., Ross, J. S., Seyler, K. L., Yao, W. \& Xu, X. Valleytronics in 2D materials, \textit{Nat. Rev. Mater.} \textbf{1}, 16055 (2016).

\bibitem{Qi:2015}Qi, J., Li, X., Niu, Q. \& Feng, J. Giant and tunable valley degeneracy splitting in MoTe$_2$, \textit{Phys. Rev. B} \textbf{92}, 121403(R) (2015).

\bibitem{Schwingenschlogl:2016} Zhang, Q., Yang, S. A., Mi, W., Cheng, Y. \& Schwingenschl\"{o}gl, U. Large Spin-Valley Polarization in Monolayer MoTe$_2$ on Top of EuO(111), \textit{Adv. Mater.} \textbf{28}, 959 (2016)

\bibitem{Scharf:PRL:2017} Scharf, B., Xu, G., Matos-Abiague, A. \& \v{Z}uti\'{c}, I. Magnetic Proximity Effects in Transition-Metal Dichalcogenides: Converting Excitons, \textit{Phys. Rev. Lett.} \textbf{119}, 127403 (2017).

\bibitem{Zollner:2019} Zollner, K., Faria Junior, P. E. \& Fabian, J. Proximity effects in MoSe$_2$ and WSe$_2$ heterostructures with CrI$_3$:  Twist angle, layer, and gate dependence.  \textit{Phys. Rev. B} \textbf{100}, 085128 (2019).

\bibitem{Zhang:2019} Zhang, Z., Ni, X., Huang, H., Hu, L. \& Liu, F. Valley splitting in the van der Waals heterostructure WSe$_2$/CrI$_3$: The role of atom superposition.  \textit{Phys. Rev. B} \textbf{99}, 115441 (2019).

\bibitem{Xie:2019} Xie J., Jia, L., Shi, H., Yang, D. \& Si, M., Electric field mediated large valley splitting in the van der Waals heterostructure WSe$_2$/CrI$_3$, \textit{Jpn. J. Appl. Phys.} \textbf{58}, 010906 (2019)

\bibitem{Lyons:NatComm:2020} Lyons, T. P., Gillard, D., Molina-S\'{a}nchez, A., Misra, A., Withers, F., Keatley, P. S., Kozikov, A., Taniguchi, T., Watanabe, K., Novoselov, K. S., Fern\'{a}ndez-Rossier, J. \& Tartakovskii, A. I. Interplay between spin proximity effect and charge-dependent exciton dynamics in MoSe$_{2}$/CrBr$_{3}$ van der Waals heterostructures, \textit{Nat. Commun.}  \textbf{11}, 6021 (2020).

\bibitem{Chen:Science:2019} Chen, W., Sun, Z., Wang, Z., Gu, L., Xu, X., Wu, S. \& Gao, C. Direct observation of van der Waals stacking–dependent interlayer magnetism, \textit{Science} \textbf{366}, 983-987 (2019).

\bibitem{Kim:PNAS:2019} Kim, H. H., Yang, B., Li, S., Jiang, S., Jin, C., Tao, Z., Nichols, G., Sfigakis, F., Zhong, S., Li, C., Tian, S., Cory, D. G., Miao, G.-X., Shan, J., Mak, K. F., Lei, H., Sun, K., Zhao, L. \& Tsen, A. W. Evolution of interlayer and intralayer magnetism in three atomically thin chromium trihalides, \textit{Proc. Natl. Acad. Sci. U.S.A.} \textbf{116}, 11131-11136 (2019).

\bibitem{Soriano:2020} Soriano, D., Katsnelson, M. I. \& Fernandez-Rossier, J. Magnetic Two-Dimensional Chromium Trihalides: A Theoretical Perspective, \textit{Nano Lett.} \textbf{20}, 6225-6234 (2020).

\bibitem{Fisher:critical:1974} Fisher, M. E. The renormalization group in the theory of critical behavior, \textit{Rev. Mod. Phys.} \textbf{46}, 597-616 (1974).

\bibitem{Stier:2016}Stier, A. V., McCreary, K. M., Jonker, B. T., Kono, J. \& Crooker, S. A. Exciton diamagnetic shifts and valley Zeeman effects in monolayer WS$_2$ and MoS$_2$ to 65 Tesla. \textit{Nat. Commun.} \textbf{7}, 10643 (2016). 

\bibitem{Tong:2020} Xu, G., Zhou, T., Scharf, B. \& \v{Z}uti\'{c}, I., Optically Probing Tunable Band Topology in Atomic Monolayers, \textit{Phys. Rev. Lett.} \textbf{125}, 157402 (2020).

\bibitem{Kim:transfer:2016} Kim, K., Yankowitz, M., Fallahazad, B., Kang, S., Movva, H. C., Huang, S., Larentis, S., Corbet, C. M., Taniguchi, T., Watanabe, K., Banerjee, S. K., LeRoy, B. J. \& Tutuc, E. van der Waals Heterostructures with High Accuracy Rotational Alignment, \textit{Nano Lett.} \textbf{16}, 1989-1995 (2016).

\bibitem{xfermatrix} Robert, C. \textit{et al.}, Optical spectroscopy of excited exciton states in MoS2 monolayers in van derWaals heterostructures, \textit{Phys. Rev. Mater.} \textbf{2}, 011001(R) (2018).

\bibitem{Kresse1999} Kresse, G. \& Joubert, D. From ultrasoft pseudopotentials to the projector augmented-wave method, \textit{Phys. Rev. B} \textbf{59}, 1758 (1999).

\bibitem{Kresse1996} Kresse, G. \& Furthm{\"u}ller, J. Efficient iterative schemes for ab initio total-energy calculations using a plane-wave basis set, \textit{Phys. Rev. B} \textbf{54}, 11169 (1996).

\bibitem{Kresse1993} Kresse, G. \& Hafner, J.J. Ab initio molecular dynamics for open-shell transition metals, \textit{Phys. Rev. B} \textbf{48}, 13115 (1993).

\bibitem{Perdew1996} Perdew, J. P., Burke, K. \& Ernzerhof, M. Generalized gradient approximation made simple, \textit{Phys. Rev. Lett.} \textbf{77}, 3865 (1996).

\bibitem{Dudarev} Dudarev, S. L., Botton, G. A., Savrasov, S. Y., Humphreys, C. J. \& Sutton, A. P. Electron-energy-loss spectra and the structural stability of nickel oxide: An LSDA+U study \textit{Phys. Rev. B} \textbf{57}, 1505 (1998).
    
\bibitem{Kormanyos2015} Korm{\'a}nyos, A., Burkard, G., Gmitra, M., Fabian, J., Z{\'o}lyomi, V., Drummond, N. D. \& Fal’ko, V. k{\textperiodcentered} p theory for two-dimensional transition metal dichalcogenide semiconductors, \textit{2D Mater.} \textbf{2}, 022001 (2015).

\bibitem{pyprocar} Herath, U. \textit{et al.} \textit{Computer Phys. Comm.} \textbf{251}, 107080 (2020).

\end{thebibliography}
\end{document}